\documentclass{jfm}
\usepackage{graphicx}
\usepackage{natbib}
\usepackage{subfigure}

\ifCUPmtlplainloaded \else
  \checkfont{eurm10}
  \iffontfound
    \IfFileExists{upmath.sty}
      {\typeout{^^JFound AMS Euler Roman fonts on the system,
                   using the 'upmath' package.^^J}%
       \usepackage{upmath}}
      {\typeout{^^JFound AMS Euler Roman fonts on the system, but you
                   dont seem to have the}%
       \typeout{'upmath' package installed. JFM.cls can take advantage
                 of these fonts,^^Jif you use 'upmath' package.^^J}%
      }
  \else
  \fi
\fi

\ifCUPmtlplainloaded \else
  \checkfont{msam10}
  \iffontfound
    \IfFileExists{amssymb.sty}
      {\typeout{^^JFound AMS Symbol fonts on the system, using the
                'amssymb' package.^^J}%
       \usepackage{amssymb}%
       \let\le=\leqslant  \let\leq=\leqslant
       \let\ge=\geqslant  \let\geq=\geqslant
      }{}
  \fi
\fi

\ifCUPmtlplainloaded \else
  \IfFileExists{amsbsy.sty}
    {\typeout{^^JFound the 'amsbsy' package on the system, using it.^^J}%
     \usepackage{amsbsy}}
    {}
\fi

\newsavebox{\astrutbox}
\sbox{\astrutbox}{\rule[-5pt]{0pt}{20pt}}

\def\pd{\partial}

\newcommand{\khat}{\ensuremath{\mathbf{\hat{k}}}}
\newcommand{\uvec}{\ensuremath{\mathbf{u}}}
\newcommand{\adv}{\ensuremath{\uvec\cdot\nabla}}
\newcommand{\nuss}{\ensuremath{\mathrm{Nu}}}
\newcommand{\pran}{\ensuremath{\mathrm{Pr}}}

\title[Dynamics of fingering convection I]{Dynamics of fingering convection I: Small-scale fluxes and large-scale instabilities}

\author[A. Traxler, S. Stellmach, P. Garaud, T. Radko and N. Brummell]%
{A. \ns T\ls R\ls A\ls X\ls L\ls E\ls R$^1$, \ns
S. \ns S\ls T\ls E\ls L\ls L\ls M\ls A\ls C\ls H$^{2,1,3}$, \break
P. \ns G\ls A\ls R\ls A\ls U\ls D$^1$,\ns
T. \ns R\ls A\ls D\ls K\ls O$^4$, \ns \and
N. \ns B\ls R\ls U\ls M\ls M\ls E\ls L\ls L$^1$ }

\affiliation{$^1$Applied Mathematics and Statistics, Baskin School of Engineering, University of California, Santa Cruz, CA 96064, USA\\[\affilskip]
$^2$Institut f\"ur Geophysik, Westf\"alische Wilhelms-Universit\"at M\"unster, D-48149 M\"unster, Germany, \\[\affilskip]
$^3$ Institute of Geophysics and Planetary Physics, University of California, Santa Cruz,CA 96064, USA\\[\affilskip] $^4$Department of Oceanography, Naval Postgraduate School, Monterey, CA 93943, USA}
\pubyear{2010}
\volume{???}
\pagerange{???--???}
\date{?? and in revised form ??}

\begin{document}
\maketitle

\begin{abstract}
 Double-diffusive instabilities are often invoked to explain enhanced
  transport in stably-stratified fluids. The most-studied natural
  manifestation of this process, fingering convection, commonly occurs in
  the ocean's thermocline and typically increases diapycnal mixing by two orders of magnitude over molecular diffusion. Fingering convection is also often associated with structures on much larger scales, such as thermohaline intrusions, gravity waves and thermohaline staircases.  In this paper, we present an exhaustive study of the phenomenon from small to large scales.  We perform the first three-dimensional simulations of the process at realistic values of the heat and salt diffusivities and provide accurate estimates of the induced turbulent transport.  Our results are consistent with oceanic field measurements of diapycnal mixing in fingering regions.  We then develop a generalized mean-field theory to study the stability of fingering systems to large-scale perturbations, using our calculated turbulent fluxes to parameterize small-scale transport.  The theory recovers the intrusive instability, the collective instability, and the $\gamma$-instability as limiting cases.  We find that the fastest-growing large-scale mode depends sensitively on the ratio of the background gradients of temperature and salinity (the density ratio).  While only intrusive modes exist at high density ratios, the collective and $\gamma$-instabilities dominate the system at the low density ratios where staircases are typically observed.  We conclude by discussing our findings in the context of staircase formation theory.
\end{abstract}


\section{Introduction}
\label{sec:intro}

When the density of a fluid depends on (at least) two components,
nominally stably-stratified systems can, under certain circumstances,
undergo double-diffusive instabilities leading to significant vertical
buoyancy transport. Here, we focus on the case of the
``fingering'' instability, which often occurs in fluids which are
thermally stably stratified, but have an inhomogeneous composition.
A well-known example is found in upper layers of the Earth's oceans
where evaporation exceeds precipitation,
leading to warm salty water overlaying colder fresh
water \citep{stern1960sfa,schmitt2005edm}.
Since heat diffuses faster than salt, parcels of
fluid displaced downward rapidly lose their heat excess
while maintaining a larger salt concentration. They become
denser than the environment and continue to sink, forming
structures called ``salt-fingers''.
Similar fingering instabilities can occur in any other
thermally stably stratified solution, provided the
concentration of the slower-diffusing solute increases
with height \citep{stern1960sfa,schmitt1983csf}.

The saturated state of this instability, fingering convection,
takes the form of tightly-packed, vertically-elongated
plumes of sinking dense fluid
and rising light fluid \citep{stern1960sfa,kunze2003ros}, and
significantly enhances the vertical transport of both heat
and chemical composition. In the ocean, fingering convection increases
diapycnal mixing within extended
regions \citep{schmitt1994ddo,kluikov1995mot,you2002goc} of the thermocline
by at least two orders of magnitude over molecular diffusion.
It has been argued that the
nutrient supply of the upper ocean \citep{dietze2004iwi}, the surface
temperature and the surface fluxes of CO$_2$ and O$_2$ are all
affected by this process \citep{glessmer2007sid}.
Conditions favorable for fingering convection also exist in many other natural
systems \citep{schmitt1983csf}. In the astrophysical context for example, a
variety of situations lead to the development of unstable mean molecular weight
gradients in otherwise stably stratified ``radiative'' regions within stars and
giant planets \citep{vauclair2004mfa,stancliffe2007cem,charbonnel2007tmp}. Since the long-term
thermal evolution and chemical stratification of these objects is regulated by
the transport bottleneck caused by radiative regions, the presence or absence
of mixing by fingering convection can influence their observable
properties dramatically.

While fingering convection is by nature a small-scale
phenomenon, it also has an intriguing propensity to
generate dynamical structures on very large scales,
such as internal gravity waves,
thermohaline intrusions and thermohaline staircases.
As first argued by
\citet{stern1969cis} and \citet{holyer1981cis}, a homogeneous field of fingers
can become unstable to the so-called ``collective instability''
leading to the spontaneous generation of
internal gravity waves in regions of active salt fingering.
This instability was later confirmed in laboratory experiments
by \citet{stern1969sfa} and in direct numerical simulations by \citet{stern2001sfu}.

Thermohaline intrusions are different
kinds of large-scale structures which are nevertheless
also often associated with fingering convection. They
take the form of laterally interleaving layers with distinct
temperature and salinity signatures, and can spontaneously form
in fluids which are stratified {\it both} vertically and horizontally.
They are commonly observed in the ocean \citep{ruddick2003obs}, %(Ruddick \& Kerr?, Ruddick \& Richards 2003),
and have been reproduced in lab experiments \citep{ruddick1979lab,ruddick1999lab}
and numerically \citep{simeonov2007}.
See \citet{ruddick2003the} for a detailed discussion of intrusion theory.

Finally, one of the most dramatic signatures of active fingering convection
in the ocean is the formation of mixed layers separated by salt finger
interfaces, known as thermohaline staircases.  Persistent staircases
have been documented in the Tyrrhenian Sea, below the Mediterranean
outflow, and in the western tropical North Atlantic \citep{schmitt1994ddo}.
Layer formation is also observed in laboratory experiments
\citep{stern1969sfa,krishnamurti2003ddt}. Layering
enhances vertical mixing by up to an order of magnitude
\citep{schmitt2005edm,veronis2007} relative to globally
similarly stratified regions characterised by a smoother stratification.
Forty years after the discovery of this phenomenon in oceanographic
field measurements \citep{tait1968sot,tait1971ts}, a generally
accepted explanation is still lacking.  \citet{radko2003mlf} argued through theoretical and numerical analyses that
the observed layering is likely to be caused by the so-called $\gamma$-instability---an instability
driven by variations in the ratio of the turbulent heat and salt fluxes.

The conventional approach to analysing the spontaneous generation of
structures from fingering convection uses the assumed
separation of scale between the finger scale and the emerging structure
scale to construct a mean-field theory, in which the
effect of the small-scale fingering is modeled through turbulent
fluxes. The resulting mean-field equations describe the evolution
of the large-scale fields only and can straightforwardly be analysed for linear stability.
Globally speaking, the modulation of the background stratification
by large-scale temperature and salinity perturbations induces
a modulation of the turbulent fluxes.
When the divergence or convergence of these
modified fluxes act to enhance the original perturbation,
large-scale modes of instability are excited.
Different variants of mean-field models have been individually successful
in representing the gross properties of each of the aforementioned
large-scale phenomena (intrusions, collective instability, $\gamma$-instability).

In this paper, we show that these various modes of instability
can actually be described by a single unifying mean-field formalism, and are
all recovered as limiting cases of our theory---each one corresponding
to a different feedback mechanism between the large-scale
perturbation and the induced turbulent fluxes. In \S \ref{sec:mft}
we present our unified mean-field model, and its relationship with
previous work. In \S \ref{sec:smallflux}, we then perform a
series of 3D simulations for parameter values typical of salty water
in the ocean, designed to measure the turbulent transport
of heat and salt as parametric functions of the background stratification.

Using the small-scale flux laws derived, we then calculate and discuss
in \S \ref{sec:discuss} the expected
growth rates of the various large-scale modes of instability
as functions of the overall stratification of the region.
Our results indicate that the relative importance of these various modes
is highly sensitive to the density ratio
(the ratio of the vertical temperature and
salinity gradients normalised by their expansion/contraction coefficients).
For low density ratio the dynamics of the system are primarily controlled by the collective
and $\gamma$-instabilities.
For intermediate density ratios,
the $\gamma$-instability is suppressed and the dynamics are
dominated by gravity waves, with intrusive modes
gaining importance. For larger values of the density ratio,
only intrusive modes are unstable.

Finally, we discuss our findings in \S \ref{sec:conclusion}, focussing on the implication of the measured turbulent fluxes for oceanic mixing in \S \ref{conclusion_fluxlaws}, and on the role of the large-scale instabilities studied in the formation of thermohaline staircases in \S \ref{conclusion_mft}.

\section{Generalised mean-field theory of fingering convection}
\label{sec:mft}

\subsection{The governing equations for homogeneous fingering convection}
\label{sec:equations}

Fingering convection, when observed in natural systems, typically occurs
far from physical boundaries. For this reason, we adopt an approach
which minimises boundary effects by considering triply-periodic
temperature, salinity and velocity perturbations driven by a steady and
uniform fingering-unstable background stratification. This
setup has been advocated by others before for studying fingering
convection \citep{stern2001sfu,radko2003mlf}, and is ideally suited
to numerical simulations using spectral methods (see \S \ref{sec:smallflux} and Paper II).
It is important to note that it does not suffer from the well-known
pathology of thermal convection in a triply-periodic system -- the so-called
homogeneous Rayleigh-B\'enard problem \citep{borueorszag1996hrb,Calzavarinietal2006hrb},
where the fastest growing modes span the entire domain and depend sensitively on the aspect ratio of the box.  Instead, the typical length scale of convective motions in the fingering regime is
set by the diffusive length scales and is independent of the box size, provided the box is large enough (see Appendix).

We consider a Cartesian coordinate system $(x,y,z)$ with $z$
increasing  upward in the vertical direction.  In all that follows, we
use the  Boussinesq approximation. We assume that the background
temperature and  salinity profiles $T_0(x,z)$ and $S_0(x,z)$ are
bilinear functions of  $x$ and $z$, $T_0(x,z)=T_{0x}x+T_{0z}z$ and
$S_0(x,z)=S_{0x}x+S_{0z}z$.   Without loss of generality, the
background fields are assumed to be two-dimensional (2D) by aligning the
horizontal gradients with the $x$-axis.  We assume \citep[as in][]{walsh1995int}
that the overall horizontal density gradient is zero, in which
case $\alpha T_{0x} = \beta S_{0x}$ where $\alpha$ and $\beta$ are the
coefficients of thermal expansion and compositional  contraction
respectively. The slope of the background temperature gradient  in
this coordinate system is $\phi=T_{0x}/T_{0z}$.

We perform a standard non-dimensionalisation procedure for studying
local fingering convection. We use the expected finger scale \citep[see][]{stern1960sfa} as the length scale,
$[l] = d = (\kappa_T\nu/g\alpha T_{0z})^{1/4}$,
where $g$ is gravity, $\nu$ is viscosity and $\kappa_T$ is thermal diffusivity.
We then define the
corresponding thermal diffusion time scale, $[t] = d^2/\kappa_T$, the
velocity scale $[u] = \kappa_T/d$, and the temperature and salinity
scales, $[T] = T_{0z}d$ and $[S] = (\alpha/\beta)T_{0z}d$.
Nondimensional parameters of interest are the
Prandtl number, $\pran = \nu/\kappa_T$, the background density ratio,
$R_0 = \alpha T_{0z}/\beta S_{0z}$ and the diffusivity ratio, $\tau =
\kappa_S/\kappa_T$.

The non-dimensional equations for the evolution of the
velocity field $\uvec=(u,v,w)$ and the temperature and
salinity perturbations $T(x,y,z,t)$ and $S(x,y,z,t)$ are then:
\begin{subequations}
\begin{eqnarray}
\frac{1}{\mathrm{Pr}} \left(\frac{\partial\uvec}{\partial t} + \adv \uvec\right) & = & -\nabla p + (T - S)\khat + \nabla^2 \uvec \label{eq:momentum}, \\
\nabla \cdot \uvec &=& 0 \label{eq:continuity},  \\
\frac{\partial T}{\partial t} + \phi u + w + \adv T & = & \nabla^2 T \label{eq:heat}, \\
\frac{\partial S}{\partial t} + \phi u + \frac{1}{R_0}w + \adv S & = & \tau \nabla^2 S \label{eq:composition},
\end{eqnarray}
\end{subequations}
where $p$ is the non-dimensional pressure perturbation from hydrostatic equilibrium and $\khat$ is the unit vector in the $z$-direction.

\subsection{Generalised mean-field theory}
\label{analytical_framework}

As discussed in \S \ref{sec:intro}, fingering convection is often
associated with the emergence of dynamical structures on scales
much larger than individual fingers.
%Taking advantage of
%scale separation, ``mean-field'' models have been developed
%to study the evolution of the system on the larger scales, while
%modelling the average effect of fingering convection through turbulent
%fluxes.
We begin by deriving a generalised set of mean-field equations, and then
study their linear stability to various large-scale modes.

\subsubsection{Mean-field equations}

As in \citet{radko2003mlf}, we are interested in the large-scale behaviour of the system
of equations (\ref{eq:momentum}-\ref{eq:composition})
when averaged over spatial/temporal scales of many fingers.
We introduce the notation $\overline{\cdots}$, where the overbar denotes
an averaging process which we assume may commute with spatial
and temporal derivatives. Let $\uvec=\bar{\uvec}+\uvec'$ and similarly
for $T$ and $S$, in which case $\overline{\mathbf{u}'}=\overline{T'}=\overline{S'}=0$.
The averaged governing equations now become
\begin{subequations}
\label{eqn_system}
\begin{eqnarray}
\frac{1}{\mathrm{Pr}} \left(\frac{\partial\overline{\uvec}}{\partial t} + \overline{\uvec}\cdot\nabla \overline{\uvec} \right) & = & -\nabla \overline{p} + (\overline{T} - \overline{S})\khat + \nabla^2 \overline{\uvec} - \frac{1}{\pran}\nabla \cdot \mathbf{R} \label{eq:mean_momentum}, \\
\frac{\partial \overline{T}}{\partial t} + \phi \overline{u} + \overline{w} + \overline{\uvec}\cdot\nabla \overline{T} & = & \nabla^2 \overline{T} - \nabla \cdot \mathbf{F}_T, \\
\frac{\partial \overline{S}}{\partial t} + \phi \overline{u} + \frac{1}{R_0}\overline{w} + \overline{\uvec}\cdot\nabla \overline{S} & = & \tau \nabla^2 \overline{S} - \nabla \cdot \mathbf{F}_S \label{eq:mean_composition},
\end{eqnarray}
\end{subequations}
where $R_{ij}=\overline{\uvec'_i\uvec'_j}$, and the turbulent fluxes are $\mathbf{F}_T=\overline{\uvec'T'}$, $\mathbf{F}_S=\overline{\uvec'S'}$.

In what follows we now drop the overbar and only refer to the
evolution of the large-scale fields $\uvec$, $T$, $S$.  As in
previous analyses, we assume that
the Reynolds stress term is small enough to neglect (a fact that is easily verified {\em a posteriori}), and additionally assume that
only the vertical component of the heat and salt fluxes are large
enough to be significant (so that
$\mathbf{F}_T \approx F_T \khat$, $\mathbf{F}_S \approx F_S \khat$).
However, we retain the diffusion terms in all three equations.
In nondimensional terms, the turbulent fluxes are characterised
by the Nusselt number $\nuss$ and the turbulent flux ratio $\gamma$, defined as:
\begin{eqnarray}
\nuss & = & \frac{F_T - (1+\pd T/\pd z)}{-(1+\pd T/\pd z)}, \label{def_Nu} \\
\gamma & = & \frac{F_T}{F_S} \label{def_gamma}.
\end{eqnarray}
Note that these definitions of $\nuss$ and $\gamma$ differ somewhat from those of \citet{radko2003mlf}, who includes the molecular diffusive terms in his definition of $F_T$ and $F_S$.  Our formalism has greater generality, since it allows for horizontal diffusive fluxes.  The difference becomes important at low Nusselt number.

We now make the key assumption that at any given time both $\nuss$ and
$\gamma$ depend only on the local value of the density ratio $R_\rho$,
which in nondimensional terms is
\begin{eqnarray}
R_\rho & = & \frac{\alpha T_{0z}(1 + \pd T/\pd z)}{\beta S_{0z}[1 + (\frac{\alpha T_{0z}}{\beta S_{0z}}) \pd S/\pd z]}, \nonumber \\
 & = & R_0 \frac{1 + \pd T/\pd z}{1 + R_0 \pd S/\pd z}. \label{def_Rrho}
\end{eqnarray}
The functions
$\nuss(R_\rho)$ and $\gamma(R_\rho)$ can be determined experimentally,
using numerical
simulations (see \S \ref{sec:smallflux}).

The system of equations describing the evolution of the large-scale quantities $\uvec$, $T$, $S$
is now
\begin{subequations}
\begin{eqnarray}
\frac{1}{\mathrm{Pr}} \left( \frac{\partial\uvec}{\partial t} + \adv\uvec\right) & = & -\nabla p + (T - S)\khat + \nabla^2 \uvec, \\
\frac{\partial T}{\partial t} + \phi u + w + \adv T & = & \nabla^2 T - \frac{\pd F_T}{\pd z}, \label{FTderiv} \\
\frac{\partial S}{\partial t} + \phi u + \frac{1}{R_0}w + \adv S & = & \tau \nabla^2 S - \frac{\pd F_S}{\pd z},  \label{FSderiv}
\end{eqnarray}
\label{avg_system}
\end{subequations}
where the flux derivative terms on the right hand side use (\ref{def_Nu}) and (\ref{def_gamma}) to express $F_T$ and $F_S$ in terms of $\nuss$ and $\gamma$.

\subsubsection{Linearised mean-field theory}
\label{sec:linmft}

The mean-field equations derived above exhibit steady solutions
describing a state of homogeneous fingering convection,
with zero mean velocity, zero deviation from the background temperature
and salinity fields %$T_0(x,z)$ and $S_0(x,z)$,
and constant
(non-dimensional) heat and salinity fluxes $F_{T0} = (1-\nuss_0(R_0))$
and $F_{S0} = \gamma(R_0) / F_{T0}$. The stability of this homogenous
turbulent state can be investigated by adding a small
perturbation to the mean quantities, and linearising the mean-field equations.
Large-scale temperature and salinity perturbations
induce large-scale variations in the density ratio,
so that $R_\rho = R_0 + R'_\rho$.
This, in turn, modulates the turbulent fluxes in a way
which may in some circumstances further enhance the initial
perturbations or quench them. Various modes
of instability are related to different feedback mechanisms between the
fields and turbulent fluxes, through the parametric functions
%$F_T(R_\rho)$ and $F_S(R_\rho)$.
$\nuss(R_\rho)$ and $\gamma(R_\rho)$.

Assuming that perturbations away from the linear background gradients are small, we first have, expanding (\ref{def_Rrho}) to linear order:
\begin{eqnarray}
R_\rho & =&  R_0\left(1 + \frac{\pd T}{\pd z} - R_0 \frac{\pd S}{\pd z}\right) = R_0 + R_\rho', \label{def_Rlin}
%\frac{\partial R_\rho}{\partial z} & = & R_0 \left(\frac{\pd^2 T}{\pd z^2} - R_0 \frac{\pd^2 S}{\pd z^2}\right)
\end{eqnarray}
which uniquely defines $R_\rho'$, and then
\begin{eqnarray}
\nuss(R_\rho) & \approx & \nuss(R_0) + \left.\frac{d\nuss}{d R_\rho}\right|_{R_0} R_\rho',
\end{eqnarray}
and similarly for $\gamma$.
Rearranging (\ref{def_Nu}) and (\ref{def_gamma}) yields $F_T=(1-\nuss)(1+\pd T/\pd z)$ and $F_S=F_T/\gamma$.  It then follows that
\begin{eqnarray}
-\frac{\partial F_T}{\partial z} & = & A_2 \left(\frac{\pd^2 T}{\pd z^2} - R_0 \frac{\pd^2 S}{\pd z^2}\right) + (\nuss_0-1) \frac{\pd^2 T}{\pd z^2}, \\
-\frac{\partial F_S}{\partial z} & = & A_1 \left(\frac{\pd^2 T}{\pd z^2} - R_0 \frac{\pd^2 S}{\pd z^2}\right)(\nuss_0-1) - \frac{1}{\gamma_0} \frac{\partial F_T}{\partial z},
\end{eqnarray}
where we have abbreviated $\mathrm{Nu}(R_0) = \mathrm{Nu}_0$, $\gamma(R_0) = \gamma_0$, and defined
\begin{eqnarray}
A_1 & = & R_0 \frac{\partial \gamma^{-1}}{\partial R_\rho}, \\
A_2 & = & R_0 \frac{\partial \mathrm{Nu}}{\partial R_\rho}.
\end{eqnarray}
Note that our $A_1$, $A_2$ are not strictly equal to those defined by \citet{radko2003mlf} but reduce to the same quantities in the limit where turbulent fluxes are much larger than diffusive fluxes.

A standard linear stability analysis of (\ref{avg_system}),
using normal modes of the form $\{\uvec, T, S\} = \{\hat{\uvec}, \hat{T}, \hat{S}\}\exp{(\lambda t + ilx + imy + ikz)}$, yields a cubic equation for the growth rate, $\lambda^3 + a_2\lambda^2 + a_1\lambda + a_0 = 0$, with
\begin{subequations}
\label{full_theory}
\begin{eqnarray}
a_2 & = & |\mathbf{k}|^2(1+\pran+\tau) + k^2\left[(1-A_1R_0)(\nuss_0-1) + A_2\left(1-\frac{R_0}{\gamma_0}\right) \right], \\
a_1 & = & |\mathbf{k}|^4(\tau\pran + \tau + \pran) + k^2|\mathbf{k}|^2 \left[(\tau + \pran)(A_2 + \nuss_0 - 1) - A_2(1+\pran)\frac{R_0}{\gamma_0} \right. \nonumber \\
 & &  - A_1R_0(1+\pran)(\nuss_0-1) \biggr] - k^4A_1R_0(\nuss_0-1)^2 + \pran\frac{l^2}{|\mathbf{k}|^2} \frac{l^2+m^2}{l^2} \left(1-\frac{1}{R_0}\right), \\
a_0 & = & |\mathbf{k}|^6\tau\pran + k^2|\mathbf{k}|^4\, \pran\left[(\tau-A_1R_0)(\nuss_0-1) + A_2\left(\tau-\frac{R_0}{\gamma_0}\right)\right] - k^4|\mathbf{k}|^2\pran R_0A_1(\nuss_0-1)^2 \nonumber \\
 & & + \pran\frac{l}{|\mathbf{k}|^2}\left\{ |\mathbf{k}|^2\left[l\, \frac{l^2+m^2}{l^2} \left(\tau - \frac{1}{R_0}\right) - k\phi(\tau-1)\right] \right. \nonumber \\
 & & + k^2A_1(1-R_0)(\nuss_0-1)(l\, \frac{l^2+m^2}{l^2} -k\phi)  \\
 & & \left. - k^2\left[A_2\left(1-R_0\right) + \nuss_0-1\right]\left[ l\, \frac{l^2+m^2}{l^2} \left(\frac{1}{R_0} - \frac{1}{\gamma_0}\right) - k\phi\left(1-\frac{1}{\gamma_0}\right)\right] \right\}, \nonumber
\end{eqnarray}
\end{subequations}
where $|\mathbf{k}|^2 = k^2 + l^2 + m^2$.

\subsubsection{Relationship with previous theories}
\label{sec:prevtheory}

Presented above is a unified formulation of several mean-field
theories, including the effects of all diffusion terms and the
contribution of variable turbulent flux ratio $\gamma$ as well as allowing for
the presence of lateral background gradients in temperature and
salinity. Several limiting cases have been discussed previously in
the literature
(unless otherwise noted these are 2D theories, so $m=0$).
\\
\\
{\bf The fingering instability. } Although technically not a mean-field
instability, it is reassuring to note that the fingering instability itself
\citep[e.g.][]{baines1969} is recovered when
turbulent fluxes and lateral gradients are ignored
($A_1=A_2=\nuss_0-1=\phi=0$). In that
case the cubic defined by (\ref{full_theory}) becomes the
well-known cubic equation for the growth rates of the fingering modes with
    \begin{subequations}
    \begin{eqnarray}
    a_2 & = & |\mathbf{k}|^2(1+\pran+\tau), \\
    a_1 & = & |\mathbf{k}|^4(\tau\pran + \tau + \pran) + \pran\frac{l^2}{|\mathbf{k}|^2}\left(1-\frac{1}{R_0}\right), \\
    a_0 & = & |\mathbf{k}|^6\tau\pran + \pran\, l^2\left(\tau-\frac{1}{R_0}\right).
    \end{eqnarray}
    \end{subequations}
\\
\\
{\bf The collective instability}, as derived by \citet{stern2001sfu},
is recovered from (\ref{full_theory}) by omitting lateral gradients ($\phi=0$),
neglecting possible variation in  $\gamma$ (so $A_1 = 0$) and
discarding the diffusion terms for temperature and salinity---but not
velocity---in (\ref{eqn_system}).
    \begin{subequations}
    \begin{eqnarray}
    a_2 & = & \pran|\mathbf{k}|^2 + k^2\left[A_2\left(1-\frac{R_0}{\gamma_0}\right) + \nuss_0-1\right], \\
    a_1 & = & \pran k^2|\mathbf{k}|^2\left[A_2\left(1-\frac{R_0}{\gamma_0}\right) + \nuss_0-1\right] + \pran\frac{l^2}{|\mathbf{k}|^2}\left(1-\frac{1}{R_0}\right), \\
    a_0 & = & \pran\frac{k^2l^2}{|\mathbf{k}|^2}[A_2(1-R_0)+\nuss_0-1]\left(\frac{1}{\gamma_0}-\frac{1}{R_0}\right).
    \end{eqnarray}
    \end{subequations}
\citet{stern1969cis} argued that modes are excited when the Stern number,
written in our notation as
\begin{equation}
A = \frac{({\rm Nu_0}-1)\left(\frac{1}{\gamma_0} - 1\right)}{{\rm Pr}\left(1-\frac{1}{R_0}\right)}, %=(\alpha_C F_C - \alpha_T F_T \;/\; (\nu \alpha_T T_{0z}  - \nu \alpha_C C_{0z})
\label{eq:stern}
\end{equation}
exceeds a value of order one. The unstable modes essentially represent overstable
gravity waves.

An elegant physical interpretation \citep{stern2001sfu} of the collective instability
is obtained by analogy with the laminar, linear double-diffusive instability
in the ``diffusive regime'', where the slowly diffusing field is stably
stratified while the rapidly diffusing field is unstably stratified.
In this case, growing oscillatory modes akin to
internal gravity waves are excited instead of fingers. Since fingering
convection induces a mean salt flux larger than the heat flux the
roles of the two fields are reversed, and the faster diffusing
field is now the salinity field. From a turbulent point
of view, the ``diffusive regime'' is recovered.
\\
\\
{\bf The theory of intrusions} of \citet{walsh1995int} is recovered by discarding the diffusion terms in (\ref{eqn_system}) and setting $\gamma$ constant ($A_1=0$), as well as neglecting Reynolds stresses in their formulation:
    \begin{subequations}
    \begin{eqnarray}
    a_2 & = & k^2\left[A_2\left(1-\frac{R_0}{\gamma_0}\right)+(\nuss_0-1)(1-A_1R_0)\right], \\
    a_1 & = & -k^4 R_0(\nuss_0-1)^2A_1 + \pran \frac{l^2}{|\mathbf{k}|^2}\left(1-\frac{1}{R_0}\right), \\
    a_0 & = & \pran \frac{lk^2}{|\mathbf{k}|^2}\left\{\left[A_2(1-R_0)+\nuss_0-1\right]\left[k\phi \left(1-\frac{1}{\gamma_0}\right) + l\left(\frac{1}{\gamma_0}-\frac{1}{R0}\right)\right] \right. \\
    & & \left. + A_1(1-R_0)(\nuss_0-1)(l-k\phi)\right\}. \nonumber
    \end{eqnarray}
    \end{subequations}
The mechanism underlying intrusive instabilities can be illustrated by imagining an alternating horizontal shear flow superimposed on the background stratification \citep[see e.g.][]{ruddick2003the}. The lack of horizontal density variation in the background (where lateral gradients of temperature and salinity compensate) implies that the background isohalines are steeper than the isothermals, and are therefore more strongly affected by horizontal advection. Alternating vertical variations in $R_\rho$ result, which in turn strengthens or weakens the fingering action, and the resulting flux convergences and divergences reinforce the intrusive motion.  Depending on the orientation of the perturbation, both direct and oscillatory modes are possible \citep{walsh1995int}.
\\
\\
{\bf The $\gamma$-instability}, as derived in \citet{radko2003mlf}, is recovered by considering horizontally invariant perturbations ($l=m=0$, $|\mathbf{k}|^2=k^2$) with zero velocity field.  From the remaining temperature and salinity equations we obtain a quadratic\footnote{To see why our formalism yields a cubic while Radko's yields a quadratic, it should be noted that (\ref{full_theory}) can be factored (with $l=m=0$) as $(\lambda
+ k^2 \pran) ( \lambda^2 + b_1 \lambda + b_0)$.  The first root
describes the viscous decay of any initial (horizontally invariant) velocity perturbation.} expression for the growth rate:
    \begin{subequations}
    \begin{eqnarray}
    a_2 & = & 1, \\
    a_1 & = & k^2\left[1 + \tau + (1-A_1R_0)(\nuss_0-1)+A_2\left(1-\frac{R_0}{\gamma_0}\right)\right], \\
    a_0 & = & k^4\left[(\tau-A_1R_0)(\nuss_0-1) - A_1R_0(\nuss_0-1)^2 + A_2\left(\tau-\frac{R_0}{\gamma_0}\right)\right].
    \end{eqnarray}
    \end{subequations}
Differences between these coefficients and those given in \citet{radko2003mlf} arise from our alternate definition of $\gamma$, but can be shown to be reduce to each other in the limit of large Nusselt number.  Note however that for $\gamma$-modes, which do not have any horizontal variation, the use of the total fluxes $F_T^{\mathrm{tot}} = F_T - (1+\pd T/\pd z)$ and $F_S^{\mathrm{tot}} = F_S - \tau(1/R_0+\pd S/\pd z)$ (originally advocated by Radko) recovers his much simpler quadratic with
    \begin{subequations}
    \begin{eqnarray}
    a_2 & = & 1, \\
    a_1 & = & k^2\left[(1-A_1^{\rm{tot}}R_0)\nuss_0+A_2\left(1-\frac{R_0}{\gamma_0^{\rm{tot}}}\right)\right], \\
    a_0 & = & -k^4 A_1^{\rm{tot}}R_0\nuss_0^2,
    \end{eqnarray}
    \end{subequations}
where $\gamma_0^{\rm{tot}} = F_T^{\rm{tot}}/F_S^{\rm{tot}}$ and $A_1^{\rm{tot}}=R_0\rm{d}(1/\gamma^{\rm{tot}})/\rm{d}R_\rho$.  As shown by Radko, a sufficient condition for the existence of a positive real root is that $A_1^{\rm{tot}}>0$, or in other words that $\gamma^{\rm{tot}}$ should be a decreasing function
of $R_\rho$. The physical interpretation
of this so-called  ``$\gamma$-instability'' is fairly
subtle, and is described in detail in the original paper \citep{radko2003mlf}.
\\
\\

\section{Turbulent flux laws}
\label{sec:smallflux}

To proceed forward and estimate growth time scales for the
various mean-field modes of instability
excited by fingering convection,
we need to determine the non-dimensional turbulent fluxes $\nuss$
and $\gamma$ as functions of the density ratio. Naturally, these depend on the
diffusivity ratios $\pran$ and $\tau$ relevant of the system
studied. Here, we choose to
focus on the case of salty water ($\pran = 7$ and $\tau = 0.01$),
as it is directly applicable to the oceanographic context.
Appendix A summarises the numerical algorithm, describes the experimental protocol for determining heat and salt fluxes, and discusses the problem of selection of the domain size to be used for these experiments.  What follows are the results of a body of simulations at different density ratios $R_\rho$.

\subsection{Typical results}

\begin{figure}
%\noindent\includegraphics[width=\linewidth]{./Visualization_Fingers_Le_100.pdf}
%\noindent\includegraphics[width=\linewidth]{./Visualization_Fingers_Le_100.epsf}
\noindent\includegraphics[width=\linewidth]{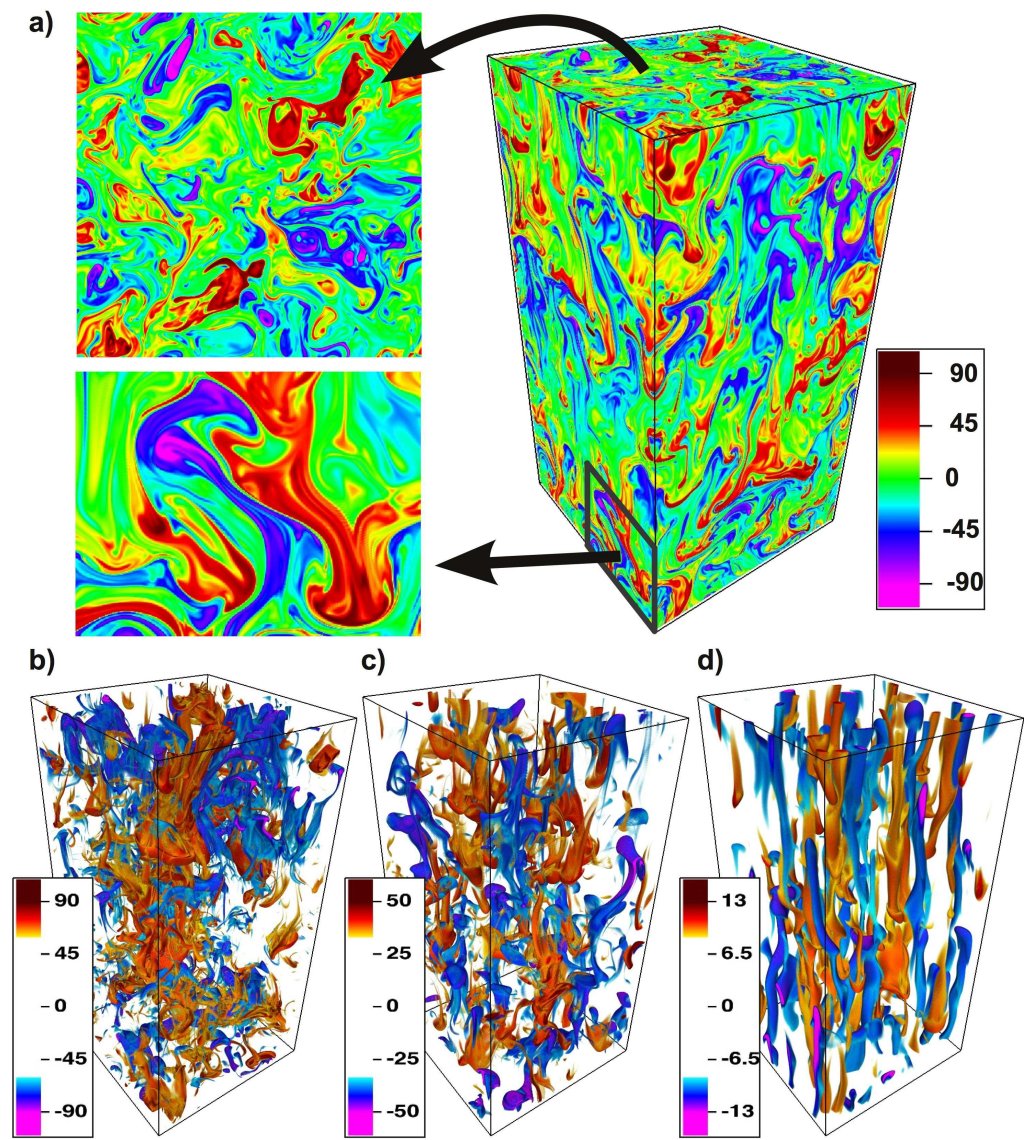}
\caption{Snapshots of the salinity field $S$ in simulations of fingering
convection in the heat-salt system ($\pran=7$, $\tau = 0.01$). {\bf a)}
Salinity field at $R_\rho=1.2$,
plotted on the three planes $x=0,y=0$ and $z=L_z$. {\bf b-d)} Volume rendering of
the salinity field for $R_\rho=1.2, R_\rho=2$ and $R_\rho=10$ (from left to right).
In all cases, the simulation domain contains $5\times 5 \times 10$ FGW (see main text).
Note how the typical amplitude of the salinity perturbation in a finger is of the
order of $1/R_\rho \tau$, or,
in dimensional terms, $d S_{oz} / \tau$. }
\label{Vis_Le_100}
\end{figure}

Figure \ref{Vis_Le_100} shows a visualisation of the salinity field obtained in the saturated state of a fingering system with $R_\rho=1.2, R_\rho=2$
and $R_\rho=10$. On account of the small diffusivity of salt compared with
all other fields, a broad range
of spatial scales exists, and a high numerical resolution is {\it a priori} required
to resolve all arising structures and correctly model the system. We find that, as expected,
the salinity field has a
complicated structure for small $R_\rho$, but successively becomes more
organised with increasing density ratio.
Regular, vertically elongated
filamentary structures dominate for $R_\rho \ge 10$.
In order to ensure that the smallest scales of the salinity field are fully resolved, we had to use the
highest resolution available (a grid of $768\times 768
\times 1536$) at $R_\rho=1.2$, although half that resolution is sufficient for $R_\rho \ge 2$.  Furthermore,
it turns out that a rough estimate of the flux laws can in fact be made with a much coarser resolution (about $32^3$).  This rather surprising result shows that the diffusion of salt does not play an important role in controlling the mixing in the heat-salt system, for very turbulent flows (low $R_\rho$).  Moreover, it suggests that low-resolution simulations may be sufficient to estimate turbulent fluxes for any high-Pr, low-$\tau$ fluid.

\begin{figure}
\begin{center}
%\noindent\includegraphics[width=0.85\linewidth]{./Flux_laws_Le_100.pdf}
%\noindent\includegraphics[width=0.85\linewidth]{./Flux_laws_Le_100.epsf}
\noindent\includegraphics[width=0.85\linewidth]{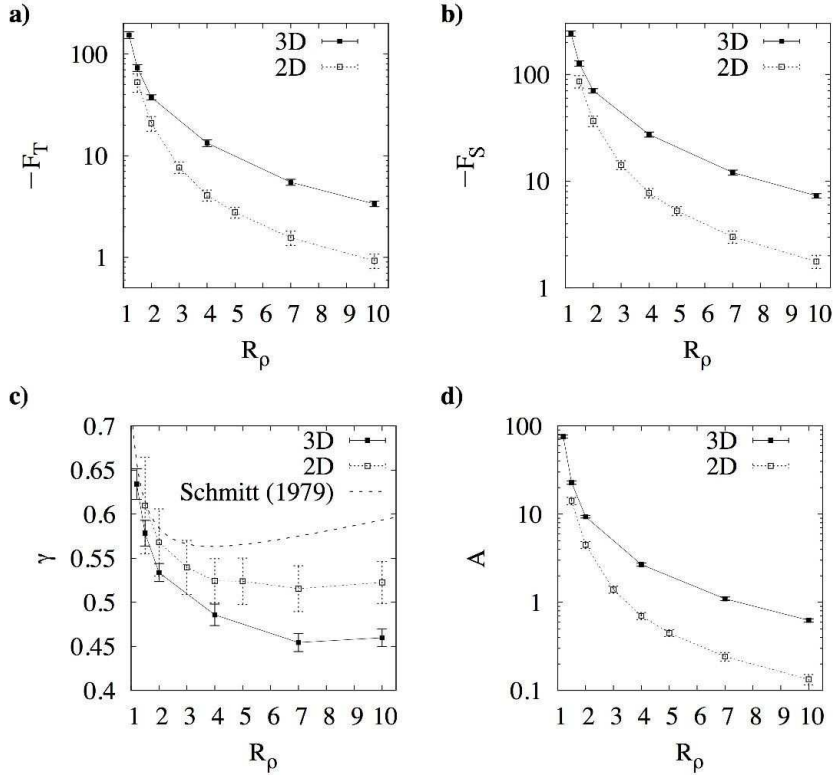}
\caption{Parametric dependence of  the non-dimensional fluxes $F_T, F_S$
as well as their ratio $\gamma$ and of the
Stern number $A$ as a function of $R_\rho$. Results from both three- and
two-dimensional simulations are shown. Panel {\bf c)} also contains a
theoretical prediction of $\gamma(R_\rho)$ based on the fastest growing
linear modes \protect\citep{schmitt1979fgm}.}
\label{Fluxes_laws_Le_100}
\end{center}
\end{figure}

\subsection{Turbulent flux laws for the heat-salt system}
\label{sec:fluxlaws}

The control parameters used in each simulation,
along with some key results, are summarised in table \ref{tab:Salt_Finger_Fluxes}.
Plots displaying the most important
findings are shown in figure \ref{Fluxes_laws_Le_100}, which also
contains results from an accompanying set of 2D simulations.

\begin{table}
  \begin{center}
\def~{\hphantom{0}}
  \begin{tabular}{l|llllll}
&  $R_\rho = 1.2$   & $ R_\rho= 1.5$    & $R_\rho=2.0$      & $R_\rho=4$         & $R_\rho=7$        & $R_\rho=10$ \\
                                                   &                                     &                                    &                                    &                                    &                                   & \\
  resolution                                 &   $768^2 \times 1536$ & $768^2 \times 1536$  & $384^2 \times   768$  & $384^2 \times   768$  & $384^2 \times   768$ & $384^2 \times   768$ \\
  $\Delta t_{\mbox{average}}$    & $39.1$                         & $57.8$                        & $121.2$                      & $223.8$                      & $422.7$                     & $390.1$ \\
  $|F_T|$                              & $153.5 \pm 11.7$        & $73.2 \pm 5.7$           & $37.6 \pm 2.2$           & $13.3 \pm 1.0 $          &  $5.48  \pm 0.38 $     & $3.35 \pm  0.21 $  \\
  $|F_S|$                              & $241.8 \pm 13.1$        & $126.4\pm 7.5$          & $70.3 \pm 3.1$           & $27.4 \pm 1.5 $          & $12.1   \pm 0.65 $     & $7.29 \pm  0.34 $  \\
  $\gamma$                                & $ 0.63  \pm  0.02$       & $0.58 \pm 0.01$         & $0.53 \pm  0.01$        & $0.49 \pm  0.01$        & $ 0.45  \pm 0.01 $     & $0.46 \pm 0.01$ \\
  $\sqrt{\left< {\bf u}^2 \right>} $
                                                   & $ 14.1 \pm  0.4$          & $9.4 \pm 0.3$             & $6.5 \pm    0.15 $       & $3.7 \pm  0.1  $          & $ 2.37 \pm    0.06$    & $1.82 \pm 0.04$ \\
  $K_T ~ [10^{-6} {\rm m}^2/{\rm s}]$          & $ 21 \pm 2$                 &  $10 \pm 1$                & $ 5.3\pm 0.3$             & $1.9 \pm 0.1   $          & $0.77 \pm 0.05$        & $ 0.47 \pm 0.03$\\
  $K_S ~ [10^{-6} {\rm m}^2/{\rm s}]$          & $ 41 \pm  2$                &  $ 27 \pm 2$               & $ 20 \pm 1$                & $15 \pm 1$                 & $11 \pm 1 $               & $ 10 \pm 0.5$     \\
  A                                               & $ 76 \pm 3 $                & $ 23 \pm 1 $               & $9.4 \pm  0.3$            & $2.7 \pm  0.1 $           & $1.1 \pm  0.05 $        & $0.63 \pm 0.03 $\\
  \end{tabular}
  \caption{Summary of simulations of fingering convection for the heat-salt system
($\tau=0.01, {\rm Pr}=7$) in a computational domain containing $5\times 5 \times 10$
fastest growing finger wavelengths \protect\citep[FGW, see][]{schmitt1979fgm}. $\Delta t_{\mbox{average}}$ denotes the length
of the time interval over which the data has been averaged, $K_T=\kappa_T |F_T|$
and $K_S=\kappa_T R_\rho |F_S| $ are the
``eddy'' diffusivities for heat and salt, taking $\kappa_T=1.4 \times 10^{-7} {\rm m}^2/{\rm s}$, and $A$ is the Stern number defined in
(\ref{eq:stern}).}
  \label{tab:Salt_Finger_Fluxes}
  \end{center}
\end{table}

As expected, we find that the turbulent fluxes $|F_S|$ and $|F_T|$ decrease
rapidly with increasing density ratio, and tend to be considerably larger
in the three-dimensional (3D) case than in 2D.
%{\bf PG: Note: Timour would like the 2D stuff removed. Comments? }
The ratio
of 3D to 2D fluxes is not constant, but tends to grow with
increasing $R_\rho$. Because of their obvious oceanic relevance, table
\ref{tab:Salt_Finger_Fluxes} also lists values for the ``eddy''
diffusivities $K_T=\kappa_T |F_T|$ and $K_S=\kappa_T R_\rho |F_S|$
of heat and salt. At $R_\rho=1.2$, we find $K_T \approx 0.21$ cm$^2$/s and
$K_S \approx 41$ cm$^2$/s. Both quantities quickly
decrease with increasing $R_\rho$.

\begin{figure}
\begin{center}
\noindent\includegraphics[width=0.5\linewidth]{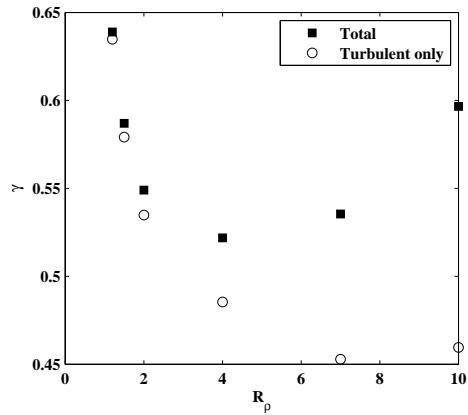}
\caption{Comparison of the flux ratio $\gamma^{\rm{tot}}$ as calculated using the total heat and salt fluxes (both turbulent and convective), and $\gamma$ as calculated using only the fluxes due to turbulent fingering convection.  The diffusive contributions become important as $R_\rho$ increases and the fingering fluxes drop, affecting not only the values of the flux ratio but also the location of the minimum of the curve.\label{full_gamma}}
\end{center}
\end{figure}

The turbulent flux ratio $\gamma$ tends to be
larger in 2D than in 3D. It initially
decreases quickly with growing $R_\rho$, attains a minimum around
$R_\rho \approx 7$ ($\gamma \approx  0.45$ in 3D) and then slowly
increases again. A widely used theoretical prediction based on
linearly fastest growing modes, originally proposed by
\citet{schmitt1979fgm}, tends to over-estimate $\gamma$ considerably,
and also predicts the minimum to occur at a smaller ($R_\rho=4$) value of
$R_\rho$.
The total flux ratio $\gamma^{\rm{tot}}$ (see \S \ref{analytical_framework}), which plays a prominent role in the $\gamma$-instability theory, is shown in figure \ref{full_gamma}.  Its value deviates significantly from the turbulent flux ratio $\gamma$ at higher values of $R_\rho$, where the Nusselt number is lower.  As a result, the position of the minimum of the curve occurs for lower $R_\rho$, thus restricting the range for which the $\gamma$-instability is expected to $R_\rho \leq 4$.  Furthermore, since the growth rate of the instability is proportional to $\rm{d}(1/\gamma^{\rm{tot}})/\rm{d}R_\rho$, we find that $\gamma$-modes should only be significant for $R_\rho \leq 2$.

%Note that the total flux ratio $\gamma_R = (1-F_T)/(\tau/R_0-F_S)$ has a similar shape, but the minimum is closer to 1, at around $R_\rho=2$, implying that the $\gamma$-instability of Radko is only expected for $R_\rho\le 2$.
Finally, we find that the Stern number $A$, which controls the
dynamics of the collective instability,  exceeds unity for
$R_\rho \le 7$ in the 3D case, while two-dimensional simulations
considerably under-estimate $A$ and therefore underestimate the range of $R_\rho$ for which the system may be unstable to these modes.

Further discussion of the implications of these simulations is deferred to \S \ref{sec:conclusion}.  For now, the above results provide the necessary data to apply the mean-field theory of \S \ref{sec:mft} to the oceanic parameter regime.

\section{Dominant modes of instability as a function of background density ratio}
\label{sec:discuss}

\begin{figure}%[hbtp]
\begin{center}
\includegraphics[height=6in]{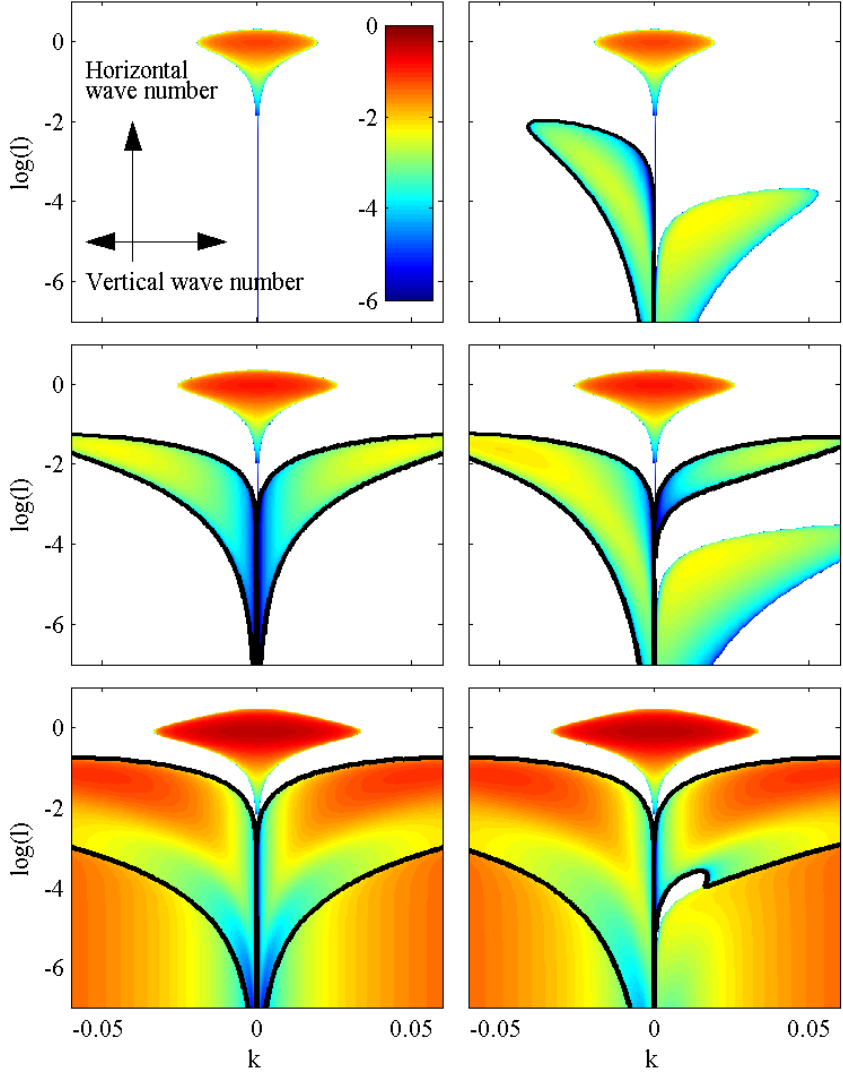}
\caption{Predicted real part of growth rates for the fastest growing perturbations, where colour is scaled to $\log{\mathrm{Re}(\lambda)}$. Only positive values are shown; the region is left white if no modes grow.
The horizontal axis shows vertical wavenumber, and the vertical axis shows the logarithm of horizontal wavenumber to capture the broad range of expected scales.  This particular display choice yields these characteristic ``flower-plots.''
The left-hand column shows results in the absence of lateral gradients ($\phi=0$) while the right-hand column shows results for a typical oceanic value of $\phi = 0.01$.  In each of the six figures, regions surrounded by a dark contour show oscillatory behaviour, by contrast with the direct modes.
For example, the symmetric ``bulbs'' at high $l$ are direct modes, corresponding to the growth rate of individual fingers.  \emph{Top:}  High density ratio ($R_0=7$).  \emph{Middle:}  Midrange density ratio ($R_0=4$).  \emph{Bottom:} Low density ratio ($R_0=1.5$).
\label{growth_colors}}
\end{center}
\end{figure}

The flux laws determined above enable
us to estimate the growth rates for the various mean-field modes of instability
discussed in \S \ref{sec:mft}. As seen in \S \ref{sec:prevtheory},
up to four modes of instability exist, but it is not immediately clear
which mode dominates in the various regions of parameter space.   To
answer this question in the oceanic context, we now examine the solutions
of the growth rate equation (\ref{full_theory}) for different values of $R_0$,
with the corresponding $\gamma_0$, $\nuss_0$, $A_1$ and $A_2$ calculated from the turbulent fluxes measured in section \ref{sec:smallflux} and shown in table 2.
We then find the largest growth rate for a given mode geometry (as determined by $k$ and $l$), maximising
$\mathrm{Re}(\lambda)$  over the three roots of the cubic.
Figure \ref{growth_colors} shows this maximum growth
rate as a function of wavenumber for three representative values of
the background density ratio:  $R_0=7$, where the
fingering instability is weak, $R_0=1.5$, where the density gradient
is close to unstable and turbulent fluxes are large (see table
\ref{partable}), and an intermediate value of $R_0=4$.  The plots show
$l$ on a logarithmic scale to capture the wide range of relevant
horizontal lengths, from the small filaments of the fingering
instability up through extensive lateral intrusions.

Since the growth rate of the fingering instability is recovered from our mean-field theory when
$A_1$, $A_2$, etc.\ are zero (as discussed in \S \ref{sec:prevtheory}), an analogous feature
appears here even though $A_1, A_2\neq 0$.  This ``fingering'' mode appears as a ``bulb'' on all plots, at the smallest horizontal scales (large $l$) and large vertical scales (low $k$).  Note, however, that mean-field theory should not be applied to model such
small-scale structures.  In practice, the bulb merely serves to indicate the region of $l$ space ($\log{l} \geq -1$) above which the theory is no longer applicable.
%Taking into account that the mean field
%approach is only valid if the averages used to derive
%(\ref{avg_system}) are taken over at
%least a few characteristic finger scales, we conclude that the
%predictions of figure \ref{growth_colors} become
%meaningless for modes with very small horizontal wavelength $\pi/l$
%or vertical wavelength $\pi/k$. Since no coherent
%large-scale patterns were observed in our simulations containing
%$5\times 5 \times 10$ FGW, we might take $L_x=5 FGW \approx 43.2$ and
%$L_z=10 FGW \approx 86.4$ as crude lower bounds on physically
%meaningful values, which are likely to underestimate the actual real
%values. For such modes, the gravity wave growth rates typically exceed
%those for the $\gamma$-modes considerably.
%It thus appears that
%gravity-wave like instabilities should dominate the large scale
%dynamics of the heat salt system, in the parameter regime where
%staircases are typically observed.

At high density ratio (top of figure \ref{growth_colors}), in the absence
of lateral gradients, only the fingering mode remains.  The
presence of a lateral gradient introduces two additional regions of
instability, one oscillatory and one direct, both confined to large vertical
and horizontal scales (i.e., small $k$, $l$).  As expected, lateral gradients break the symmetry of the solutions, since the $k\phi$ terms in (\ref{full_theory}) distinguish between positive and negative $k$ perturbations.
The direct mode corresponds to a lateral
intrusion which typically grows on a time scale of about 30 hours, with a horizontal scale of the order of a kilometer and vertical scale of a few metres ({\it i.e.}, with a slope of the order of $\phi$).

For an intermediate
density ratio ($R_0=4$, middle panels of the figure), the gravity waves of
the collective instability
appear at a range of vertical and horizontal scales starting at $l=0.055$, $k=0.06$ (a physical size of about a metre in each direction), with a growth time scale of about 30 hours.
The lateral
gradient strongly modifies the gravity waves, increasing both their maximum growth rate and the size of the instability region for negative $k$ values, while suppressing growth for positive $k$.  As with $R_0=7$, the lateral gradient also triggers a direct intrusive mode at large horizontal scales.

Finally, at low density ratios the system is
dominated by the collective instability (oscillatory) and the
$\gamma$- (direct) instability, and is now unstable to a continuous range of
modes on both large and small horizontal and vertical scales.
At the scales for which the mean-field theory is valid ($k \ll 1$), the collective instability grows fastest, most unstable on scales of a metre both horizontally and vertically, and with a growth time scale around two hours regardless of the presence of a lateral gradient ($\phi=0.01$).  For comparison, a $\gamma$-mode of the same vertical scale grows at roughly a third this rate.

\begin{table}
\caption{Empirically derived parameters for growth rate prediction using the unified theory at high, midrange, and low values of background density $R_0$.
$A_1$ and $A_2$ are estimated by using neighboring values of the density ratio.\label{partable}}
\begin{center}
\begin{tabular}{l l l l l}
$R_0$   & $\nuss_0$ & $\gamma_0$    & $A_1$ & $A_2$ \\ \hline
7.0     & 5.48      & 0.45         & 0.16  & -11.61 \\
4.0     & 13.3      & 0.49         & 0.27  & -25.7 \\
1.5     & 73.2      & 0.58         & 0.56  & -217.31
\end{tabular}
\end{center}
\end{table}

\begin{figure}
\begin{center}
\includegraphics[height=2.5in]{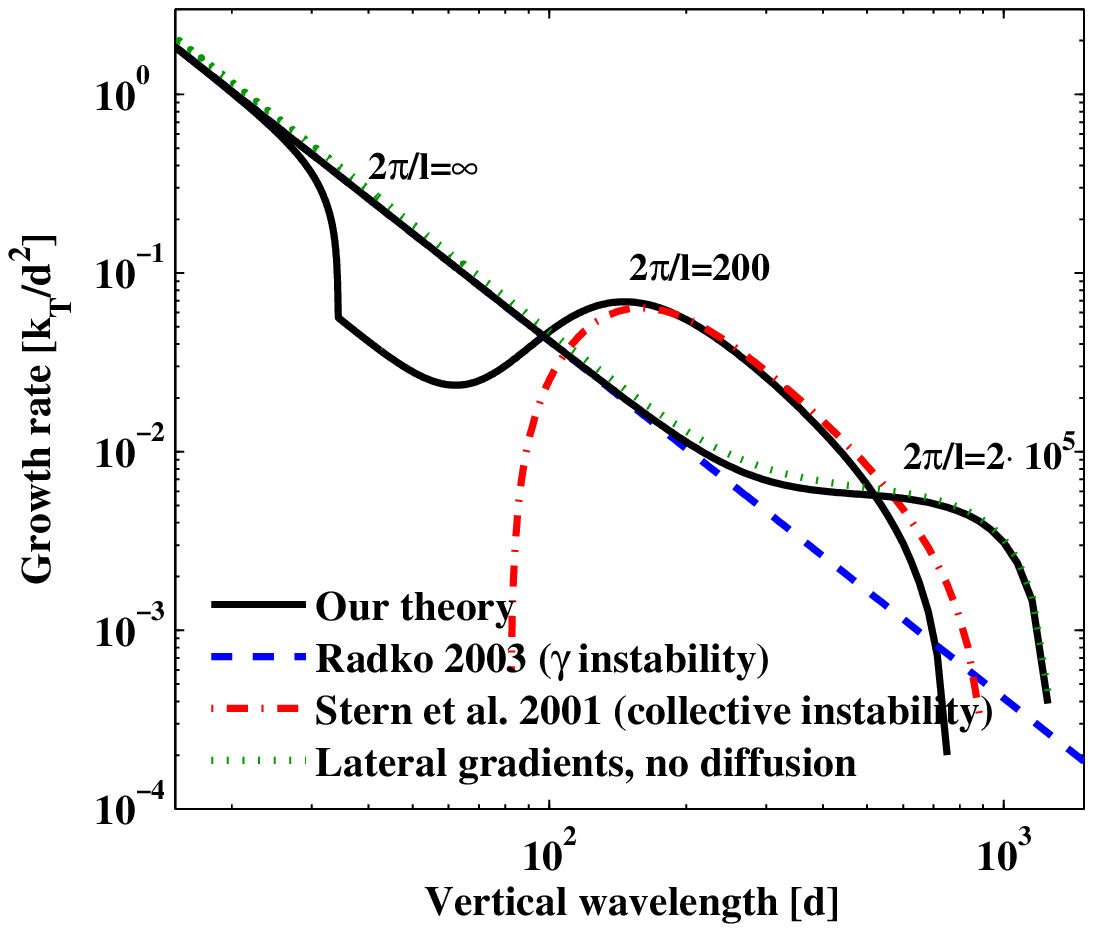}
\caption{Comparison of the real component of projected growth rates between theories at $\pran=7$, $\tau=1/100$, for $R_0=1.5$ and $\phi=0.01$.
The $l=0$ ($2\pi/l=\infty$) mode growth rate is identical to the one predicted by the $\gamma$-instability of \protect\citet{radko2003mlf}; the remaining lines have horizontal perturbation wavelengths of $200d=1.8 \mathrm{\ m}$ and $200000d=1.8 \mathrm{\ km}$.
The mean-field theory captures both the collective instability at large vertical scales and the $\gamma$-instability at small vertical scales, and matches well with the intrusive instability at the largest scales.\label{compare_polynomials}}
\end{center}
\end{figure}

Figure \ref{compare_polynomials} shows the growth rate of various modes in our theory at low density ratio ($R_\rho=1.5$) and in the presence of lateral gradients ($\phi=0.01$).
Also shown are the growth rates of the collective instability of \citet{stern2001sfu}, of the $\gamma$-instability of \citet{radko2003mlf}, and of the intrusive instability of \citet[absent Reynolds stresses]{walsh1995int}.
For a horizontally invariant perturbation ($l=0$, equivalent to a domain width of \mbox{$2\pi/l\rightarrow\infty$}), the mean-field theory matches Radko's $\gamma$-instability.  For perturbations with moderate horizontal scales ($2\pi/l=200d$, approximately two metres in dimensional terms), the unified theory predicts the presence of gravity waves with similar vertical scales and recovers the collective instability.
As the vertical wavelength decreases, the mode gets flatter, and becomes a $\gamma$-mode.
At the largest horizontal scales ($2\pi/l=2\cdot 10^5d$, about two kilometers), our theory recovers the growth rate of intrusive modes from \citet{walsh1995int}.

Using this unified formalism, we have therefore demonstrated
how the dominant type of large-scale instability in salt
fingering systems depends strongly, not only on the scale of
the perturbations considered, but also on the background density
ratio.  Field observations \citep{you2002goc} reveal that $R_\rho$ can vary
significantly in the ocean.  In nearly unstable regions
($R_\rho \rightarrow 1$), the collective and $\gamma$-instabilities
control the dynamics of the system (also see Paper II), but as the system becomes more strongly stratified, these modes disappear and
lateral gradients emerge as the dominant factor in the creation of
large-scale structures.

\section{Discussion and Conclusions}
\label{sec:conclusion}

\subsection{Turbulent flux laws}
\label{conclusion_fluxlaws}
The simulations of \S \ref{sec:fluxlaws} are of direct relevance to the problem of
parameterising double-diffusive mixing in the ocean.  These problems
arise in basin-scale ocean circulation models
\citep{gargett1992dif,zhang1998mod,merryfield1999mod} and in fine-scale
studies focusing on the dynamics of intrusions, internal waves, and
thermohaline staircases \citep{walsh2000,simeonov2004,stern2002,radko2005dtl}.
While several attempts
have already been made to deduce the small domain flux laws from
numerical simulations \citep{shen1995,stern2001sfu,stern2005,radko2008ddm},
the computational
restrictions in early studies precluded direct treatment of the
oceanographic case (characterised by three-dimensional dynamics,
Pr=7, $\tau=0.01$).  Instead, simulations were either two-dimensional or
employed diffusivity ratios significantly higher than the heat-salt
value of 0.01.  The double-diffusive flux laws were deduced by
extrapolation of the numerical results obtained in the computationally
accessible regime---an approach clearly requiring a posteriori
validation.  The simulations summarised in table \ref{tab:Salt_Finger_Fluxes} are the first DNS that meet the
challenge of solving the actual heat/salt problem in three dimensions.
A comparison of these results with earlier studies reveals a good qualitative agreement
with earlier estimates.  In retrospect, it is perhaps surprising to see
how well the former educated guesses of flux laws \citep{schmitt1979fgm,stern2001sfu}
captured the pattern of heat/salt diffusivities as a
function of density ratio.

The comparison with oceanographic field measurements is more
ambiguous since small-scale mixing in the ocean is driven by a
combination of double-diffusion and turbulence---their relative
contribution is uncertain and much debated.  Nevertheless, the careful
analysis of the NATRE (North Atlantic Tracer Release Experiment) data
set by \citet{stlaurent1999} made it possible to evaluate the
salt finger diffusivities directly from observations. The estimated salt
diffusivity is characterised by a monotonically decreasing dependence
on density ratio, reducing from $K_S=50\cdot 10^{-6} \; {\rm m}^2/{\rm s}$
at $R_\rho=1.4$ to $K_S = 10\cdot 10^{-6} \; {\rm m}^2/{\rm s}$ at $R_\rho=1.8$
\citep{stlaurent1999}.  Once again, these
values are in broad agreement with the DNS results summarised in table 1.

Finally, the presented synthetic data (table 1) make it possible to
assess the relevance of several hypotheses proposed to explain the
physics of equilibration.  Most notably, \citet{stern1969cis} suggested that the
amplitude of salt fingers could be limited by the collective
instability, with equilibrium fluxes characterised by
values of $A\sim 1$.  A similar suggestion was put forward by \citet{kunze1987},
who pointed out that Stern's \citeyearpar{stern1969cis} criterion is equivalent to
specifying the Richardson number based on scales of individual fingers
and speculated that an increase in $A$ above unity would be followed by
the rapid destruction of fingers by secondary instabilities---an idea
most recently revisited by \citet{inoue2008}.
Our results,
which reveal variation in $A$ by two orders of magnitude, emphasise the
limitations of Stern/Kunze hypothesis and motivate the search for
alternative conceptual models.

\subsection{Large-scale instabilities and implications for the formation of thermohaline staircases}
\label{conclusion_mft}

In this paper, we have treated three proposed mechanisms for large-scale instability in salt fingering systems, unifying under one framework what had previously been studied in isolation \citep{walsh1995int,stern2001sfu,radko2003mlf}.  Note that a related approach to the problem, considering the effect of the Richardson number on the fluxes and focussing on thermohaline interleaving rather than the oscillatory modes of the collective instability, may be found in the recent work of \citet{smyth2010}.  In our work, as in theirs, considering all instability mechanisms in a single formalism opens the possibility of comparing the growth rates of the various mean-field modes to one another and establishing the dominant ones in each region of parameter space.  Furthermore, using the realistic flux laws discussed above, we are now able to give robust quantitative predictions for the presence and growth rates of each mode individually.

For the heat-salt system, we find that no single one of the proposed instability mechanisms is expected to dominate in all fingering-unstable regions of the ocean.  At high density ratio ($R_\rho \geq 7$), for example, the Stern number drops below one and only lateral intrusions may be destabilised.  As shown in \S \ref{sec:discuss}, intrusive modes with horizontal scales on the order of a kilometer and vertical scales of a few metres are expected to grow on a time scale of about 30 hours.
%Interleaving is recognized as an
%important mechanism for the cross-frontal exchange of heat and salt,
%with lateral interleaving diffusivities comparable with those caused
%by geostrophic eddies \citep{ruddick2003obs}.  \citet{garrett1982}
%suggested that interleaving can be an essential component of
%the global cascade of temperature and salinity variances to small scales,
%providing the link between lateral stirring by mesoscale variability and
%diapycnal mixing.  \citet{ruddick1988med} reinforced this view by
%demonstrating that thermohaline interleaving was responsible for the
%decay and eventual destruction of Meddy Sharon, one of the vortices
%of Mediterranean origin in the Atlantic.
As the density ratio decreases below seven, gravity-wave modes are also destabilised, growing fastest at horizontal and vertical scales of a few metres.  The relative growth rates of the gravity waves and the intrusive modes depend sensitively on their spatial extent and on the slope of the isothermal contours ($\phi$) with respect to the vertical, in a manner which can be evaluated through our theory.  Finally, when the background stratification is close to neutral stability, which is the case for most fingering regions of the ocean, $\gamma$-modes are also unstable and grow on similar time scales as the gravity waves---on the order of a few hours, much faster than the intrusive modes.

These findings also enable us to place constraints on existing theories for the formation of thermohaline staircases.  Indeed, all three mean-field modes of instability have been proposed to generate these structures in the process of their nonlinear development \citep{walsh1995int,stern2001sfu,radko2003mlf}.  However, it is important to note that staircases are typically only observed to exist in very low density ratio environments \citep[$R_\rho < 2$, see][]{schmitt1981}.  We find that in this parameter regime (see figure \ref{compare_polynomials}), intrusions grow one to two orders of magnitude slower than gravity waves or $\gamma$-modes unless lateral gradients are exceptionally strong (which could happen in some regions of the Mediterranean outflow, for example).  This limits the relevance of intrusive modes when applied to the formation of staircases in the bulk of the thermocline.  We also find that gravity-wave modes grow faster than $\gamma$-modes on all spatial scales for which mean-field theory is applicable (vertical wavelength greater than about $100d$, see Appendix).  This should {\em a priori} point to the collective instability as the mechanism responsible for layer formation.

However, the aforementioned correspondence between the locations of observed oceanic staircases ($R_\rho < 2$) and interval of strongly decreasing $\gamma^{\rm{tot}}(R_\rho)$ is too remarkable to be dismissed.  In addition, the only existing numerical simulation to date for which staircase formation has been observed \citep{radko2003mlf} has unambiguously identified a $\gamma$-mode as the staircase precursor.  For these reasons, the $\gamma$-instability could prove to be just as important as a generating mechanism for these large-scale structures.  At this point, large-scale numerical simulations are the only avenue towards further progress, an avenue we follow in part II of this paper.

%With the predicted behavior of the system so strongly dependent on the background density ratio, a complete theory is essential to fully understand the linear instability of the full range of possible fingering systems, as no one instability mechanism is expected to dominate in all regions of interest.  Finally, this analysis raises other issues beyond the question of how well its predictions will agree with experiment.  If instabilities do grow at the expected scales and rates, do they saturate through nonlinear effects not captured in this theory, or persist to create large-scale structures?  What effect will such structures have on the vertical transport of heat and salt?  Part II of this paper seeks to address these questions, focusing on the collective and $\gamma$-instabilities.

\begin{acknowledgments}
A.T., P.G.  and T.R. are supported by the National Science Foundation,
NSF-093379 and NSF-0807672, and T.R. is supported by an NSF CAREER. S.S. was supported
by grants from the
NASA Solar and Heliospheric Program (NNG05GG69G, NNG06GD44G, NNX07A2749). The simulations
were run on the Pleiades supercomputer at UCSC, purchased using an
NSF-MRI grant. Computing time was also provided by the John von Neumann
Institute for Computing. We thank Gary Glatzmaier for many helpful discussions and
for his continuous support.
\end{acknowledgments}
\appendix

\section{Numerical determination of flux laws}

\subsection{Description of the numerical algorithm}

We measure the nondimensional turbulent fluxes of heat and salt as functions of the density ratio using the following numerical algorithm.
We solve the original set of
equations (\ref{eq:momentum}--\ref{eq:composition}) for homogeneous fingering convection
using, as explained in \S \ref{sec:equations}, triply-periodic boundary conditions
for all perturbations, e.g.
\begin{equation}
T(x,y,z,t) = T(x+L_x,y,z,t) = T(x,y+L_y,z,t) = T(x,y,z+L_z,t) \mbox{  ,}
\end{equation}
where $(L_x,L_y,L_z)$ defines the dimensions of the computational box (in units of $d$).
Note that in these units the global Rayleigh number of a simulation
is equal to $L_z^4$.
In this section, all quantities refer to the full field containing all scales
(by contrast with \S \ref{analytical_framework}).
%In the case where horizontal gradients are neglected,
%note that the horizontal averages of the temperature and salinity fields
%are not assumed nor forced to be
%zero, so that the mean vertical temperature and compositional profiles can
%evolve freely with time. Another way of interpreting this triply-periodic system is that
%the point-wise difference $\Delta T = T(x,y,L_z,t) - T(x,y,0,t)$ (and similarly for $\Delta S$)
%of temperature across the height $L_z$ of the box is fixed and equal to $L_z T_{0z}$,
%although the specific values of $T(x,y,0,t)$ (say) is not. A similar argument applies
%if horizontal gradients are present.

We use a spectral algorithm based on the classical
Patterson-Orzag method \citep{canuto2007sme} widely used for simulations
of homogenous turbulence. Nonlinear products are evaluated on a grid in
physical space, and the 3/2-rule is used to avoid aliasing errors
(for reference, $N$ grid points in a coordinate direction
corresponds to Fourier modes up to wavenumbers of $(2/3)N$ in
that direction). Note that our simulation is a Direct Numerical Simulation
(DNS), with no subgrid scale model.
A third order, semi-implicit Adams-Bashforth / Backward-Differencing
algorithm \citep{peyret2002smi} is used for time-stepping.
All diffusive terms are treated implicitly, while the advection terms are
explicitly treated. The above time-stepping method was chosen since it offers
a relatively large stability domain that includes a part of the imaginary axis
at a comparatively low cost. In order to guarantee numerical stability,
the time step is adjusted dynamically. The code was designed to run efficiently
on massively-parallel supercomputers and employs a transpose-based parallel
transform algorithm \citep{stellmach2008esm}.

\subsection{Experimental protocol for determination of fluxes}

We are primarily
interested in the high-Rayleigh number limit, where for
fixed fluid parameters Pr and $\tau$ the transport
properties depend only on the background density ratio $R_0$.
This implies the need to use a reasonably tall computational box ($L_z \gg 1$ in units of $d$).
Our simulations must also permit a large enough number of
fingers to exist in the
horizontal directions to provide good statistical estimates  of the
turbulent fluxes. On the other hand, the domain size should be small
enough to suppress any secondary large-scale instabilities, which
would drive the local density ratio $R_\rho$ away from the background $R_0$
and modulate the turbulent fluxes we are trying to measure.
After a careful study of the
outcome of a series of simulations, further detailed below, we found
that a computational domain of size $5\times 5 \times 10$ in units of
the ``fastest growing wavelength'' (FGW)  is a good compromise.  The
FGW is defined as the wavelength of the fastest growing mode of the
fingering instability, and depends on the model parameters Pr, $\tau$
and $R_0$ (see e.g. figure 4 of \citet{schmitt1979fgm}). We find that it is also
a good measure of the width of the fully nonlinear fingers in the
parameter regime considered  (see figure \ref{finger_sizes} and discussion below). Finally, note that since
large-scale perturbations cannot grow in these ``small-box'' simulations,
$R_\rho = R_0$ everywhere in the domain. In this section, we will equivalently
use the notation $R_0$ or $R_\rho$ to denote the density ratio since they are the same.

\begin{figure}
%\noindent\includegraphics[width=\linewidth]{./Fluxes_Le_100.pdf}
%\noindent\includegraphics[width=\linewidth]{./Fluxes_Le_100.epsf}
\noindent\includegraphics[width=\linewidth]{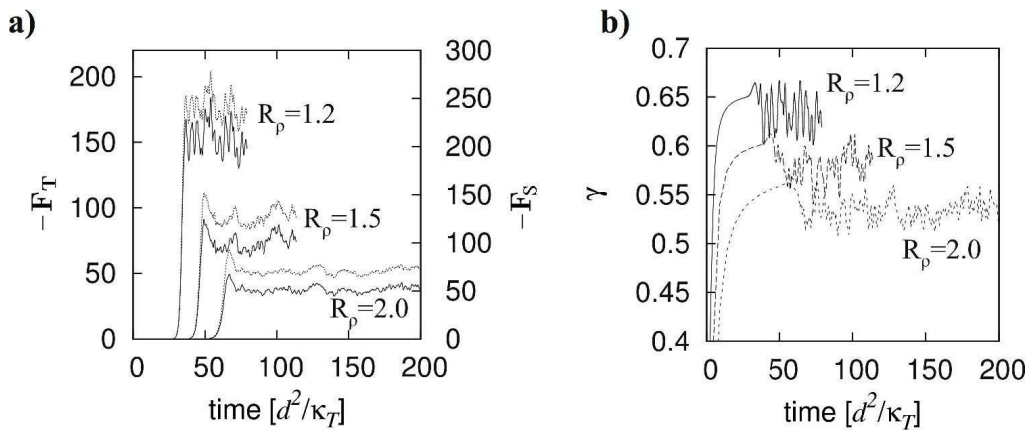}
\caption{Time series derived from simulations of fingering convection
in the heat-salt system ($\pran=7$, $\tau=0.01$) for the three cases $R_\rho=1.2$,
$R_\rho=1.5$ and $R_\rho=2$ (the first and third also shown in figure \ref{Vis_Le_100}). {\bf a)}
Fluxes $-F_T$ and $-F_S$ (note that for salt fingering $F_T$ and $F_S$ are both negative quantities), {\bf b)} Turbulent flux ratio $\gamma$. }
\label{Fluxes_Le_100}
\end{figure}

After selecting a value of the density ratio, calculating the corresponding
domain size in units of $d$,
and determining an appropriate spectral resolution for the simulations
(see below, and table \ref{tab:Salt_Finger_Fluxes}),
we initialise the calculation with low-amplitude white noise perturbations
in $T$ and $S$, and let the system evolve with time.
Vertical turbulent fluxes of heat and salt, defined as
\begin{eqnarray}
F_T=<w T>,  \nonumber \\
F_S=<w S>,
\end{eqnarray}
are then computed, where $<...>$ denotes a
volume average over the computational domain. Figure
\ref{Fluxes_Le_100} shows time series of $-F_T$ and
$-F_S$ (note that both fluxes are negative for fingering convection),
as well as that of their ratio $\gamma$, for three values of $R_\rho$ and for a
fluid with the characteristic properties of salty water ($\tau=0.01,
\pran =7$). After a period of exponential growth, the system settles
into a statistically stationary finite amplitude state in which all
plotted quantities fluctuate about well-defined temporal averages,
the state of homogeneous fingering convection.
The turbulent fluxes reported in table \ref{tab:Salt_Finger_Fluxes} are measured from temporal means of $F_T$ and $F_S$ once the system is in that state.

\subsection{The effect of the domain size}
\label{sec: flux_laws_protocol}

\begin{figure}%[htbp]
\begin{center}
% Color figure
\includegraphics[height=6in]{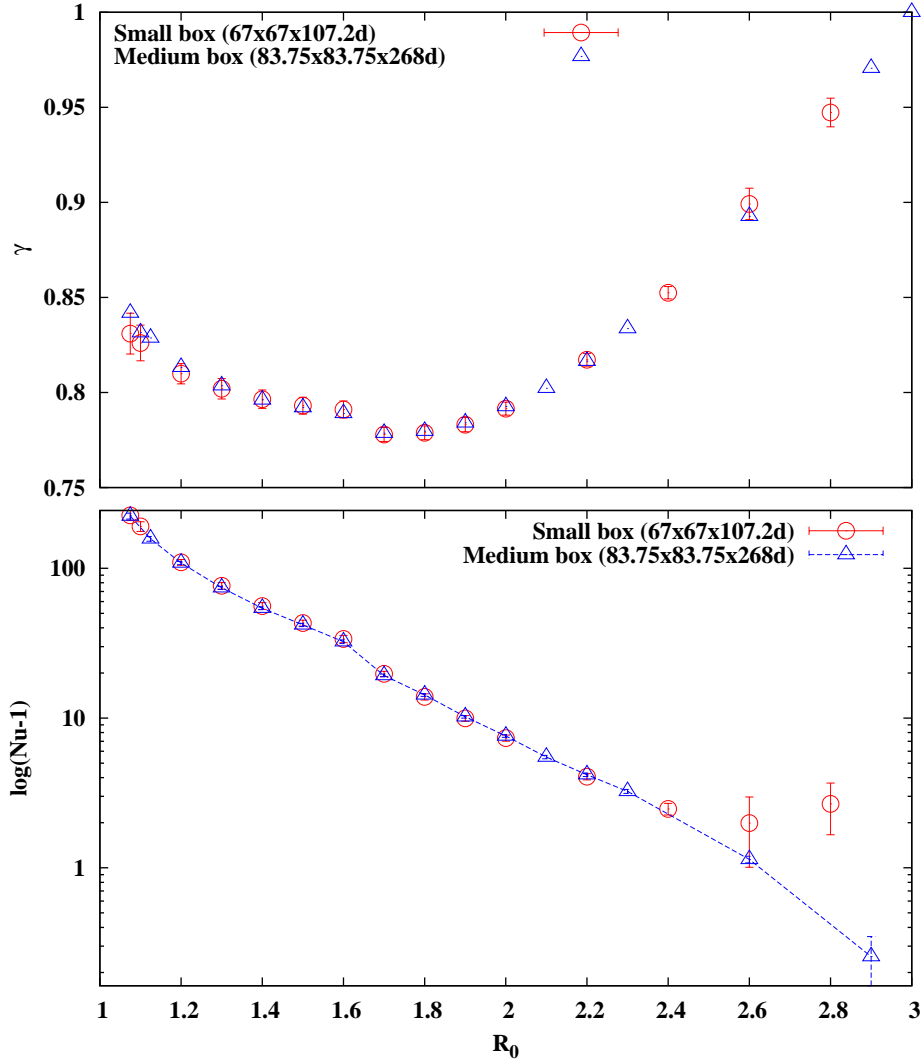}
% B&W figure
%\includegraphics[height=6in]{fig7.eps}
\caption{Small and medium box average values of the turbulent flux ratio $\gamma(R_0)$ and the logarithm of the Nusselt number $\nuss(R_0)$.  For most $R_0$ values the calculated small-box averages of $\nuss$ and $\gamma$ closely compare with their medium-box counterparts, even where the finger height exceeds the small box height (identifiable in the bottom panel of figure \ref{finger_sizes}).  At large $R_0$, however, the averages diverge, seen in the logarithmic $(\nuss_0-1)$ values.\label{gamma_Nu_compare}}
\end{center}
\end{figure}

In the last part of this appendix, we estimate the optimal box size for calculating local flux laws,
while meeting the criteria described above (large enough to provide good statistics, small enough to suppress secondary instabilities).
%Three considerations guide our choice: the box size must be large enough to ensure that it does not
%influence the dynamics of the fingers, it must also be large enough to provide good statistics when measuring the
%turbulent fluxes, and finally, it must yet be small enough to prevent the growth of the large-scale modes of
%instability which would otherwise modulate the density ratio.
Guidance for our choice can come from
simulations in the less computationally demanding parameter regime of $\pran=7$, $\tau=1/3$.
Figure \ref{gamma_Nu_compare} shows
the averaged Nusselt number and $\gamma$ values for simulations from two domains at these parameters.
The first uses a computational domain of approximately $5\times 5\times 8$ fastest-growing wavelengths (FGW), and the second is much taller, approximately $6\times 6\times 20$ FGW.
Comparison of the
turbulent fluxes as a function of $R_\rho$
%, as estimated in the two sets of simulations,
shows strong agreement between the two sets of simulations except at the highest and lowest values of the density ratio, where some divergence occurs
(discussed below).  Overall, these results suggest that the smallest box-size provides sufficient statistics and a large-enough domain to yield accurate flux laws at most density ratios. Meanwhile, neither set of simulations seem to exhibit any large-scale modes of instability, which also satisfies our requirements.

In order to understand the difference between the turbulent fluxes measured in the two geometries at large density ratio, we examine the geometry of the nonlinear fingers in more detail.
For an experimental estimate of the height of fingers in a given simulation,
we look to the autocorrelation of the vertical velocity, averaged over all $x$ and $y$:
\[
f(s) = \int\int\int_V w(x,y,z) w(x,y,z+s)\, dx\, dy\, dz
\]
We estimate a typical finger height by the distance $s$ beyond which the autocorrelation function drops below 0.05, and similarly for the typical finger width.

\begin{figure}%[hbtp]
\begin{center}
% Color figure
\includegraphics[height=3.5in]{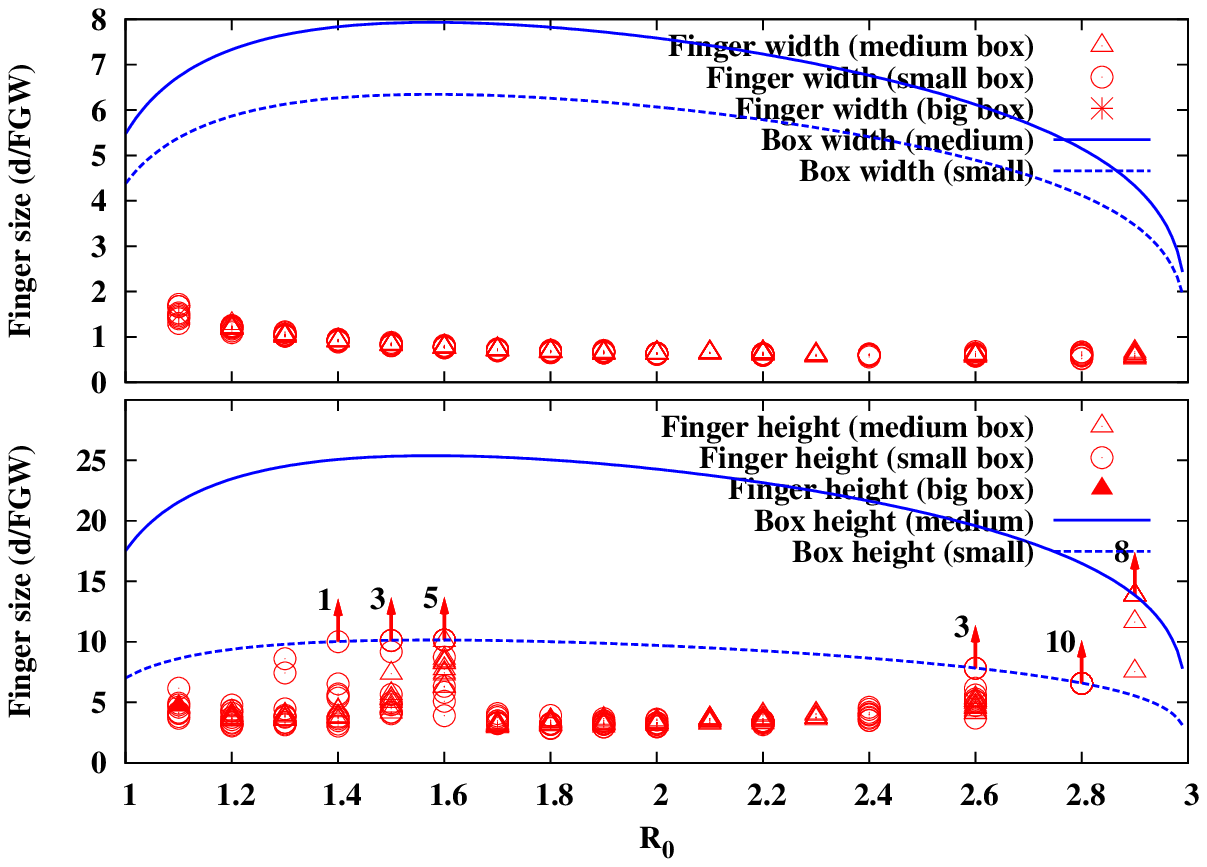}
% B&W figure
%\includegraphics[height=3.5in]{fig8.eps}
\caption{Numerically determined finger sizes for $\pran=7$ and $\tau=1/3$ over a range of density ratios at two box sizes:  small box, $67d\times 67d\times 107.2d$, and medium box, $83.75d\times 83.75d\times 268d$.  The single data point available from the S09 simulations is included for comparison.  \emph{Top:}  Finger width, demonstrating that both small and medium boxes are wide enough to contain many fingers in the horizontal dimension.  In both boxes, finger widths are uniformly of the order of one FGW.
\emph{Bottom:}  Finger height, showing the effect of decreasing box size.  Points located on the box height line represent sample times at which vertical velocity autocorrelation does not drop below the threshold value (i.e., fingers are taller than the box size); the corresponding numeric labels indicate the number of sampled points that exceeded the box height.  In the small box overshoot occurs in two regions, $R_0=1.4$--$1.6$ and $R_0\ge2.6$.  In the medium box, fingers only reach the box height for $R_0\ge 2.9$.\label{finger_sizes}}
\end{center}
\end{figure}

\begin{figure}%[hbtp]
\begin{center}
\includegraphics[height=3in]{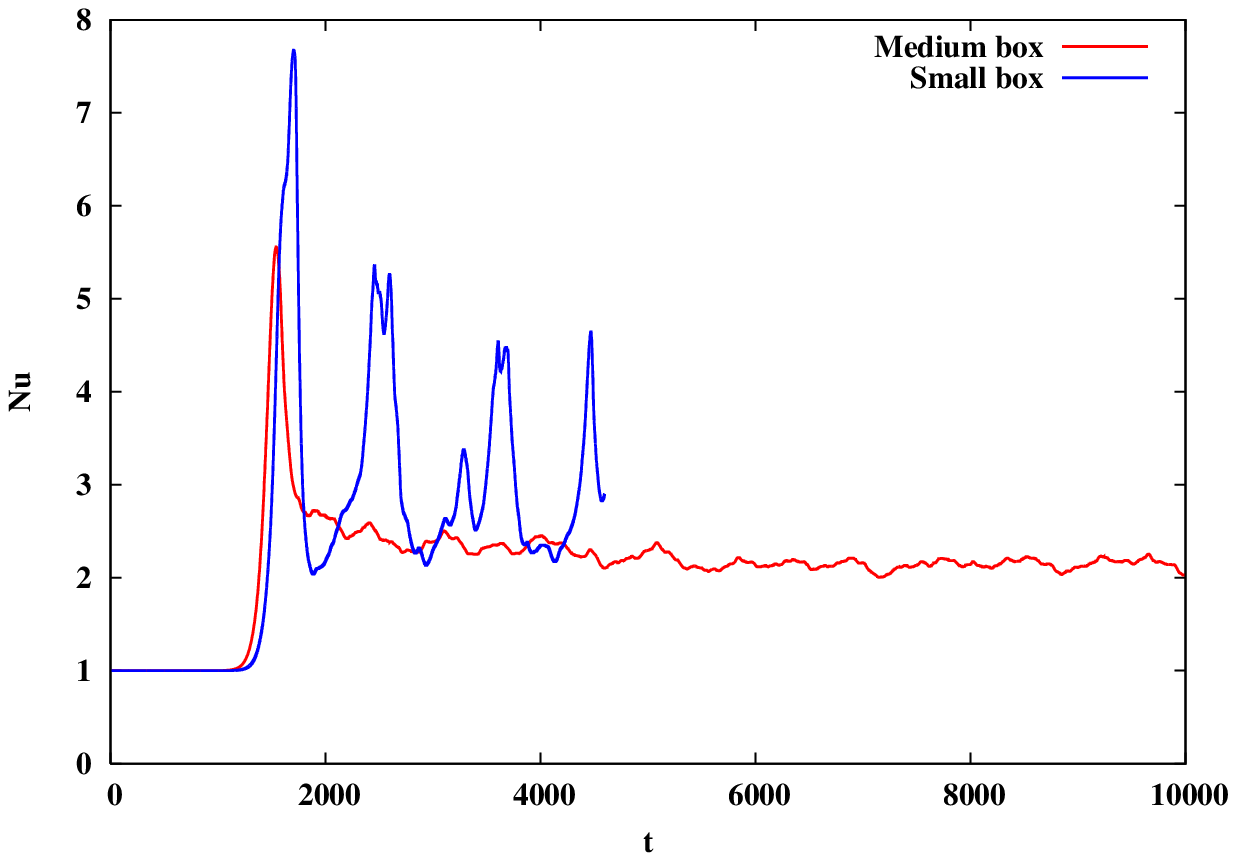}
\caption{Nusselt number time series for the medium and small boxes at $R_0=2.6$, where fingers overshoot the small box height but are still contained in the medium box height (see figure \ref{finger_sizes}).  In the medium box, the Nusselt number quickly reaches a stable average with only small fluctuations, where the small box series varies widely.\label{NuR26compare}}
\end{center}
\end{figure}

Figure \ref{finger_sizes} shows the results of these calculations for the small ($5\times 5\times 8$ FGW) and medium ($6\times 6\times 20$ FGW) domains, using ten sample points separated by 5000 time steps at each value of the background density ratio $R_0$.
By inspection of the results for the medium domain, we find that fingers are typically about 0.5--2 FGW in width, and 4--6 FGW in height, for most values of the density ratio except very close to marginal stability. As a result,
both box sizes are sufficiently wide to accommodate many fingers.  The small box is tall enough to contain typically two fingers, while the larger box contains approximately five.  Both domains seem to be satisfactory except at the largest values of $R_0$ (where elevator modes are expected to dominate as $R_0\rightarrow 1/\tau$), for which the small box is too short to contain even a single finger.  Large fluctuations in the Nusselt number result, shown in figure \ref{NuR26compare}.

For the purpose of extending mean-field theory to more difficult parameter ranges (such as $\pran \ll 1$, $\tau \ll 1$ of interest in the astrophysical context), these results provide a valuable guide.  For most values of $R_0$, boxes of no more than $5\times 5\times 10$ FGW in height may be expected to provide robust flux averages, while suppressing large-scale structures such as gravity waves that would otherwise swamp the finger field and complicate the averaging process.  However, as the background density ratio increases toward a completely stable stratification, taller boxes are necessary to prevent finger overshoot from artificially increasing the measured fluxes.

One final feature bears comment, namely the sharp increase and sudden decrease in finger height variability as $R_0$ increases above 1.6.  This transition, apparently not a function of box size, corresponds to a kink in the $\gamma(R_\rho)$ curve (see figure \ref{gamma_Nu_compare}), but speculation as to its cause is deferred to future work.

\bibliographystyle{jfm}
\bibliography{DDC_bib}

\end{document}